\documentclass[aps,prb,superscriptaddress,showpacs,twocolumn]{revtex4}
\usepackage{epsfig}
\begin{document}
\title{Ruthenocuprates RuSr$_{2}$(Eu,Ce)$_{2}$Cu$_{2}$O$_{10-y}$:
Intrinsic
magnetic multilayers}
\author{I.\v Zivkovi\'c}
\affiliation{Institute of Physics, P.O.B.304, HR-10 000, Zagreb,
Croatia}
\author{Y.Hirai}
\affiliation{Physics Department, University of Wisconsin, Madison,
WI 53706,
U.S.A.}
\author{B.H.Frazer}
\affiliation{Physics Department, University of Wisconsin, Madison,
WI 53706,
U.S.A.}
\author{M.Prester}
  \email{prester@ifs.hr}
\affiliation{Institute of Physics, P.O.B.304, HR-10 000, Zagreb,
Croatia}
 \author{D.Drobac}
\affiliation{Institute of Physics, P.O.B.304, HR-10 000, Zagreb,
Croatia}
\author{D.Ariosa}
\affiliation{Institut de Physique Appliqu\'ee, \'Ecole Polytechnique
F\'ed\'erale de Lausanne, CH-1012 Lausanne, Switzerland}
\author{H.Berger}
\affiliation{Institut de Physique Appliqu\'ee, \'Ecole Polytechnique
F\'ed\'erale de Lausanne, CH-1012 Lausanne, Switzerland}
\author{D.Pavuna}
\affiliation{Institut de Physique Appliqu\'ee, \'Ecole Polytechnique
F\'ed\'erale de Lausanne, CH-1012 Lausanne, Switzerland}
\author{G.Margaritondo}
\affiliation{Institut de Physique Appliqu\'ee, \'Ecole Polytechnique
F\'ed\'erale de Lausanne, CH-1012 Lausanne, Switzerland}
\author{I.Felner}
\affiliation{Racah Institute of Physics, Hebrew University,
Jerusalem, Israel}
\author{M.Onellion}
  \email{onellion@landau.physics.wisc.edu}
\affiliation{Physics Department, University of Wisconsin, Madison,
WI 53706,
U.S.A.}
\date{\today}
\begin{abstract}
{We report ac susceptibility data on
RuSr$_{2}$(Eu,Ce)$_{2}$Cu$_{2}$O$_{10-y}$ (Ru-1222, Ce content x=0.5
and 1.0), RuSr$_{2}$GdCu$_{2}$O$_{8}$ (Ru-1212) and SrRuO$_{3}$. Both
Ru-1222 (x=0.5, 1.0) sample types exhibit unexpected magnetic dynamics
in low magnetic fields: logarithmic time relaxation, switching
behavior, and `inverted' hysteresis loops. Neither Ru-1212 nor
SrRuO$_{3}$ exhibit such magnetic dynamics. The results are
interpreted as evidence of the complex magnetic order in Ru-1222. We
propose a specific multilayer model to explain the data, and note that
superconductivity in the ruthenocuprate is compatible with both the
presence and absence of the magnetic dynamics.} 
\end{abstract}
\pacs{74.27.Jt, 74.25.Ha, 75.60.Lr, 75.70.Cn}
\maketitle

\section{Introduction}

Coexistence of superconductivity and long-range magnetic order, and
the types of magnetic order compatible with superconductivity, are
problems of widespread interest~\cite{map}. Such systems include the
ruthenocuprates, which exhibit superconductivity in the CuO$_{2}$
planes~\cite{tok} with some type(s) of long-range magnetic order that
involves at least the RuO$_{2}$ planes~\cite{fel1,ber}. One of the
main issues for the ruthenocuprates is the nature of long-range
magnetic order coexisting with superconductivity.  The issue is
complicated because, as previous work (muon spin rotation~\cite{ber},
magnetic resonance~\cite{fei}, and neutron diffraction~\cite{lyn}) has
shown, there is evidence- even in the simple Ru-1212 material- for
both ferromagnetic~\cite{ber,fei} and antiferromagnetic~\cite{lyn}
ordering. Both magnetization~\cite{wil} and NMR~\cite{tok} studies of
Ru-1212 materials confirm the presence of a ferromagnetic component of
the low temperature magnetic order.  Theoretical
calculations~\cite{nak} of the electronic structure predict
antiferromagnetic order for Ru-1212.

There is an implicit assumption that all ruthenocuprates will possess
the same long-range magnetic order. As we show below, this is not the
case for the data we measured, comparing Ru-1212 and Ru-1222, nor is
it the case for the existing literature.  Neutron diffraction
measurements have not been reported for Ru-1222. Magnetization, low
frequency susceptibility, and M\"ossbauer/NQR reports~\cite{fel1,wil1}
indicate a pronounced- perhaps even a dominant- role of ferromagnetism
in the spontaneous magnetic order of Ru-1222.  The main result of our
report is that Ru-1222 samples exhibit unexpected dynamical magnetic
features in very low magnetic fields.  We have measured the AC
susceptibility while varying the dc magnetic field either continuously
or in steps. We found a pronounced susceptibility `switching',
logarithmic time relaxation, and hysteretic, inverted-in-sense,
susceptibility butterfly loops.  While these properties have been
individually reported earlier in other, non-superconducting magnetic
systems, the Ru-1222 system exhibits all of these properties.  The
inverted butterfly hysteresis represents, to our knowledge, the first
observation of this phenomenon in a bulk magnetic system.  The
contrast between Ru-1222 and Ru-1212 (or SrRuO$_{3}$) samples is
marked: neither Ru-1212 nor SrRuO$_{3}$ exhibit any of these dynamical
magnetic properties.  Following the data, we present a model in which
we argue that the magnetic ordering of Ru-1222 involves both
ferromagnetic and antiferromagnetic coupling, and that Ru-1222 is a
rare example of an intrinsic, naturally grown magnetic multilayer
system with the layers coupled with antiferromagnetic interactions,
similar to that inferred of the (La,Sr)$_{3}$Mn$_{2}$O$_{7}$ colossal
magnetoresistance manganite~\cite{wel}.
\begin{figure}[t]
\epsfig{file=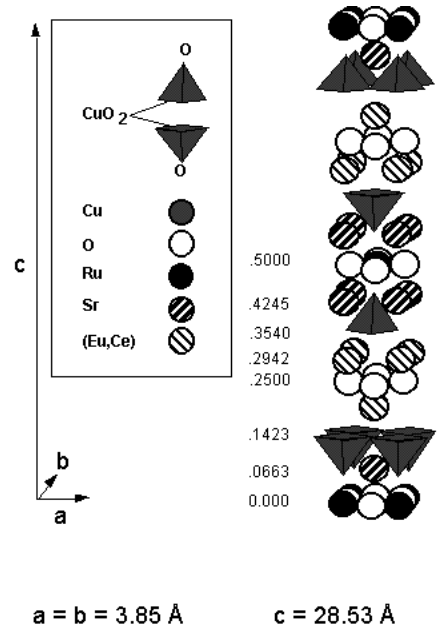,width=5.8cm,clip=}
\caption{Unit cell of Ru-1222 ruthenocuprate}
\label{f1}
\end{figure}

\section{Experimental}

Polycrystalline samples of Ru-1222, Ru-1212 and SrRuO$_{3}$ were
fabricated as published elsewhere~\cite{fel1,berg}.  Two Eu/Ce
stoichiometries of Ru-1222 were synthesized: 'superconducting' (Ce
content x= 0.5) and 'insulating' (Ce content x=1.0).  SrRuO$_{3}$
served as a three dimensional, ferromagnetic reference material.  In
all three materials, magnetic order stems from the RuO$_{6}$
octahedra.  We used x-ray diffraction (data not shown) to establish
that all samples were crystallographically single phase.  We measured
the sample microstructure using scanning electron microscopy (SEM),
with results shown in Figure 2.  AC susceptibility data were taken
using a CryoBIND system~\cite{cry} calibrated for absolute
susceptibility results. The AC susceptibility measurements used a
frequency of 230 Hz, an ac magnetic field of 0.15 Oe, and a dc
magnetic field between 0-100 Oe. DC susceptibility measurements of
some samples were obtained using a Lake Shore vibrating-sample
magnetometer (VSM).

\section{Results}

Figure 1 illustrates the Ru-1222 unit cell.  Notice the large
separation along the c-axis between RuO$_{2}$ planes; we return to
this point below. Figure 2 illustrates the microstructure.
%
\begin{figure}[t]
\epsfig{file=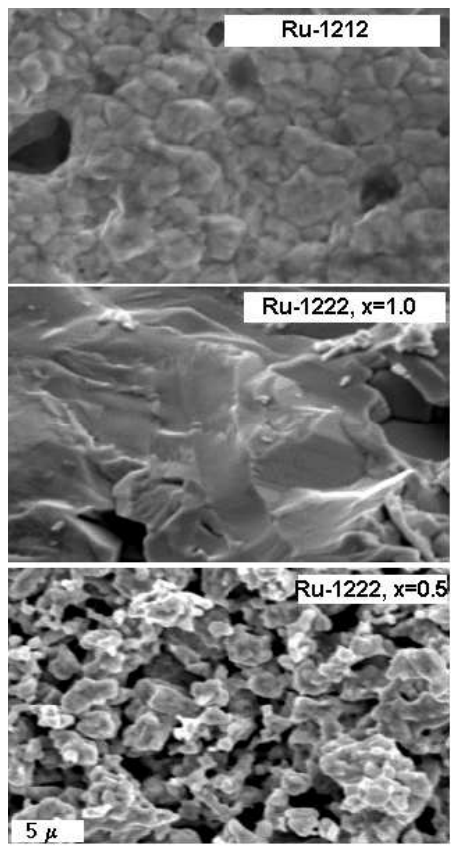,width=6cm,clip=}
\caption{Scanning electron microscopy images of Ru-1212 (top
panel), Ru-1222, x=1.0 (middle panel), and Ru-1222, x=0.5 (bottom
panel)}
\label{f2}
\end{figure}
Ru-1222 (x=1.0) samples exhibit a dense structure with almost no
isolated grains and very small intergranular regions; grain
boundaries were
difficult to identify. Ru-1222 (x=0.5) and Ru-1212 samples, by
contrast, exhibit well-defined grains (size typically $1-2 \mu m$)
and pronounced grain boundaries.  As we note further below, there are
marked, qualitative differences between the Ru-1212 samples and Ru-
1222 samples of either stoichiometry. Fig.\ \ref{f2} is thus important
because it rules out grain structure as the source of these
qualitative differences.

Figure 3 shows the AC susceptibility data of Ru-1212, Ru-1222 (x=0.5)
and Ru-1222 (x=1.0). Ru-1212 exhibits a single maximum at T$_{N}$=133
K. By contrast, both Ru-1222 samples exhibit peaks at lower
temperature T$_{M}$ (= 85 K (x=0.5) and 117 K (x=1.0)) and a broad
feature between 120-140 K, followed by non-Curie-Weiss behavior
extending up to 180K.  We discuss possible interpretations of the
broad features and non-Curie-Weiss behavior separately~\cite{ari}.

In this work we primarily focus the magnetically ordered
(T$<$T$_{M}$) phase of both 
ruthenocuprates and report hysteretic and highly nonlinear magnetic
dynamics 
characterizing Ru-1222, but not Ru-1212, samples.
\begin{figure}[b]
\epsfig{file=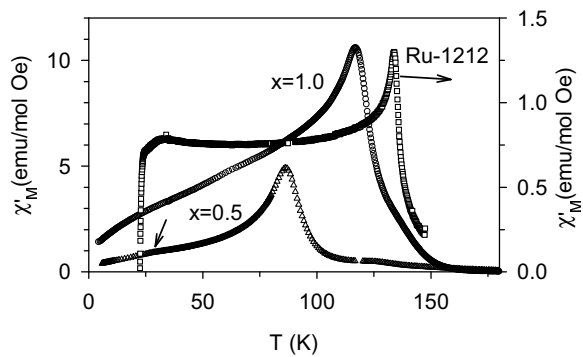,width=7.8cm,clip=}
\caption{Ac susceptibility measurements of Ru-1222 (x=0.5 and x=1.0)
and Ru-1212 samples. Note different scales for the two sample types.
Vertical arrow indicates the kink attributed to superconductivity in
Ru-1222, x=0.5, sample.}
\label{f3}
\end{figure}
Figures 4-10 illustrate different aspects of the AC susceptibility
response of Ru-1222, pointing out the differences in the equivalent
response
of Ru-1212 under similar experimental conditions.
\begin{figure}[t]
\epsfig{file=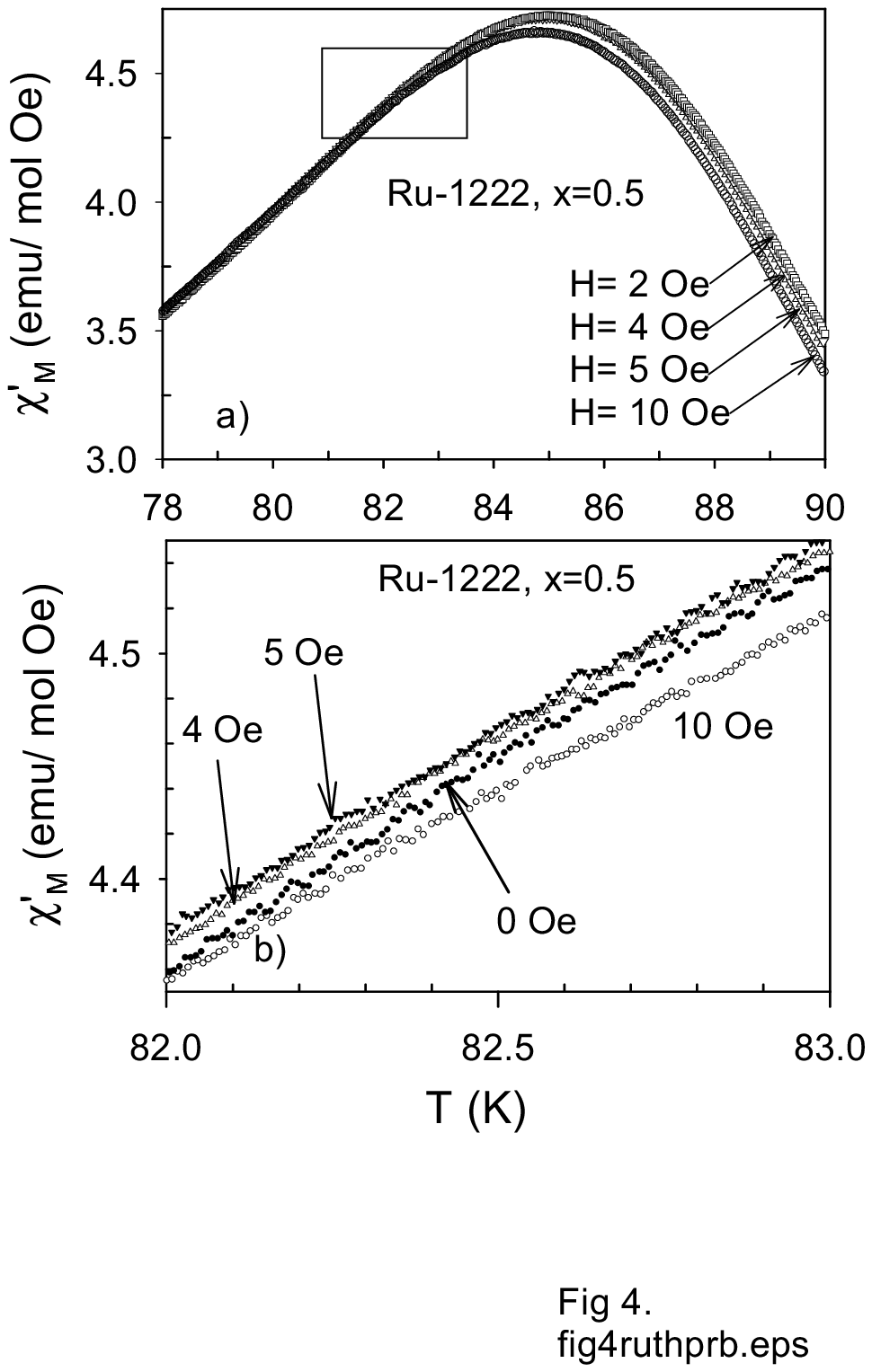,width=7cm,clip=}
\caption{Temperature dependence of ZFC ac susceptibility of Ru-1222,
x=0.5, sample in a sequence of small applied magnetic fields. The
panel a) would suggest collapsing of all of the curves below the
ordering maximum inside the 10 Oe magnetic field range. A closer
inspection of the rectangular area, shown in b) on expanded scale,
indicates that below the ordering maxima there is actually a non-
monotonic change of ac susceptibility as the applied magnetic field
increases.}
\label{f4}
\end{figure}
Although we show only the results for x=0.5 or x=1.0 compositions of
the Ru-1222 ruthenocuprate in a particular figure, both
stoichiometries exhibit the same qualitative behavior in all respects.
We first report on AC susceptibility temperature dependence in small
($<$ 100 Oe) applied dc magnetic fields. Figs.\ \ref{f4},\ \ref{f5}
illustrates a sequence of temperature dependences of zero-field cooled
(ZFC) AC susceptibility measurement in several dc magnetic fields at
temperatures near the peak value for Ru-1222 (x=0.5 and x=1.0)
samples. These figures show that there is a range of dc magnetic field
values, and temperatures, for which the AC susceptibility of Ru-1222
samples increases as the dc magnetic field increases.  By contrast,
Ru-1212 and SrRuO$_{3}$ samples exhibit `normal' behavior: at all
temperatures, the AC susceptibility monotonically decreases for
increasing dc magnetic fields. Qualitatively, `normal' behavior is
easy to interpret: AC susceptibility measures how free the magnetic
moments are to perform forced oscillations imposed by the ac magnetic
field. Therefore, any superimposed dc magnetic field introduces a
further restriction on the oscillations, and the AC susceptibility
decreases. Fig.\ \ref{f5} illustrates AC susceptibility for Ru-1222
(x=1.0) samples. Note the common qualitative trend shown in Fig.\
\ref{f4}: below T$_{M}$ the AC susceptibility versus dc magnetic field
exhibits non-monotonic behavior with increasing dc magnetic field,
while for temperatures above T$_{M}$  the AC susceptibility decreases
monotonically with increasing dc magnetic field.The unexpected
increase in AC susceptibility from zero dc field to the `turning
field' (= the dc field at which the AC susceptibility is a maximum)
reaches as much as 15\% for Ru-1222 (x=1.0).  It is much less
($\approx$ 0.5\%) for Ru-1222 (x=0.5), and the magnitude of the
turning field is lower for Ru-1222 (x=0.5) than for Ru-1222 (x=1.0).
\begin{figure}[t]
\epsfig{file=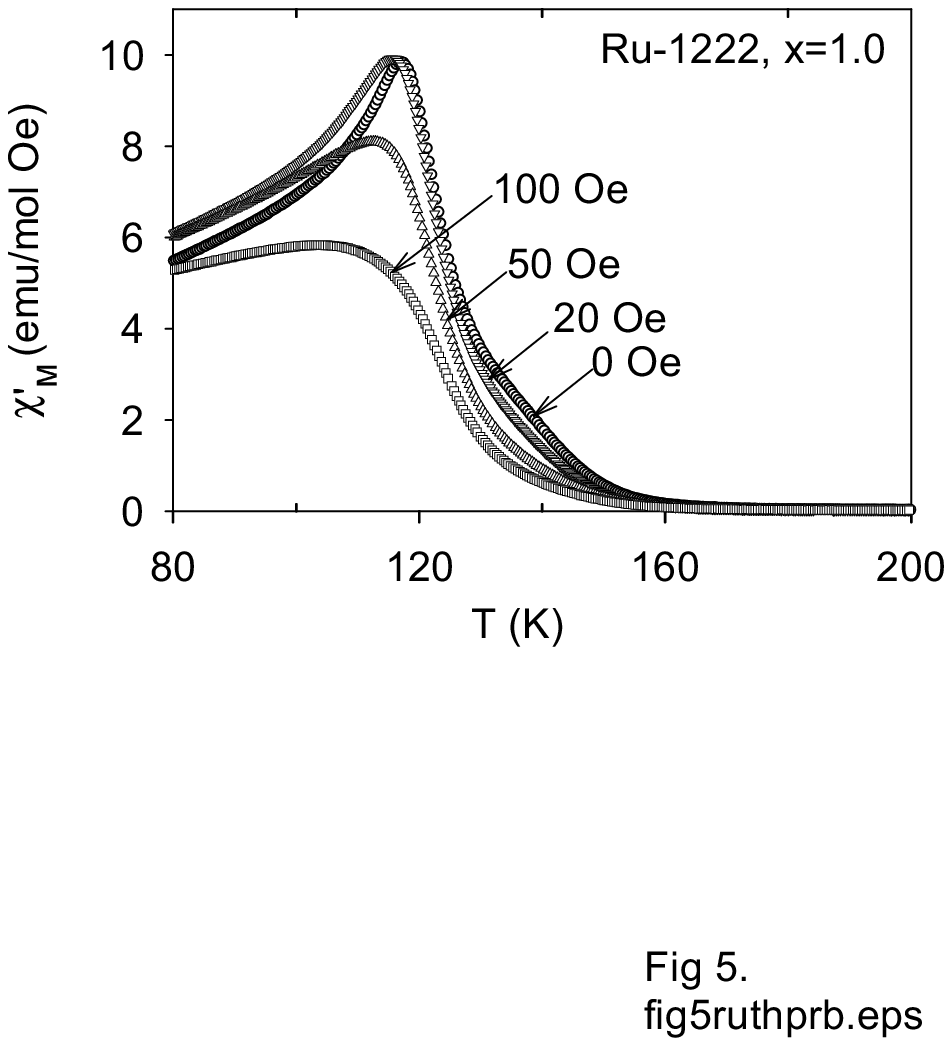,width=7.5cm,clip=}
\caption{Temperature dependence of ZFC ac susceptibility of Ru-1222,
x=1.0, sample versus applied dc magnetic field.  Below the ordering
maxima, the susceptibility exhibits non-monotonic behavior versus
applied magnetic field.}
\label{f5}
\end{figure}
For some samples we also noted small, quasi-periodic jumps in the
temperature
dependence of the AC susceptibility below T$_{M}$, similar to the
jumps recently
reported for certain manganite samples~\cite{mur1}. Because the
occurrence of these oscillations/jumps were not reproducible in
repeated measurements,
we did not perform any systematic studies of this effect on our
samples.

\subsection{Time dependence of AC susceptibility}

We measured the time response of the AC susceptibility to different dc
magnetic fields at several fixed temperatures.  We would zero-field
cool the sample to a fixed temperature, apply a dc field and take the
measurement of time dependence, then- in zero dc magnetic field- raise
the temperature to above 180K and lower the temperature (to the same
or another fixed temperature) before repeating the measurement with
different dc field. Fig.\ \ref{f6} illustrates some of the results.
The inset of Fig.\ \ref{f6}a shows how the dc magnetic field was
abruptly turned on and off. Figs.\ \ref{f6}a) and b) shows that the AC
susceptibility sharply increaseses (`switches') when the dc magnetic
field is turned on. After the AC susceptibility switch, there is a
gradual decrease.  In Figs.\ \ref{f6}a) and b), note that even after
2000 sec. the AC susceptibility has not returned to the initial value.
The AC susceptibility exhibits a strong time relaxation. Such
relaxation -often called disaccommodation~\cite{kro,mur}- has been
reported previously for other magnetic systems~\cite{kro}. The data in
Figs.\ \ref{f6}a) and b) are, however, surprising in certain respects:
the AC susceptibility {\em increases} above the ZFC value when a dc
magnetic field is applied.  As Fig.\ \ref{f6}c shows, the AC
susceptibility relaxes slowly, and logarithmically, and does not
return to the ZFC value on the time scale of at least one day (the
longest period we measured at one dc field and temperature). The data
indicate:
\begin{figure}[ht]
\epsfig{file=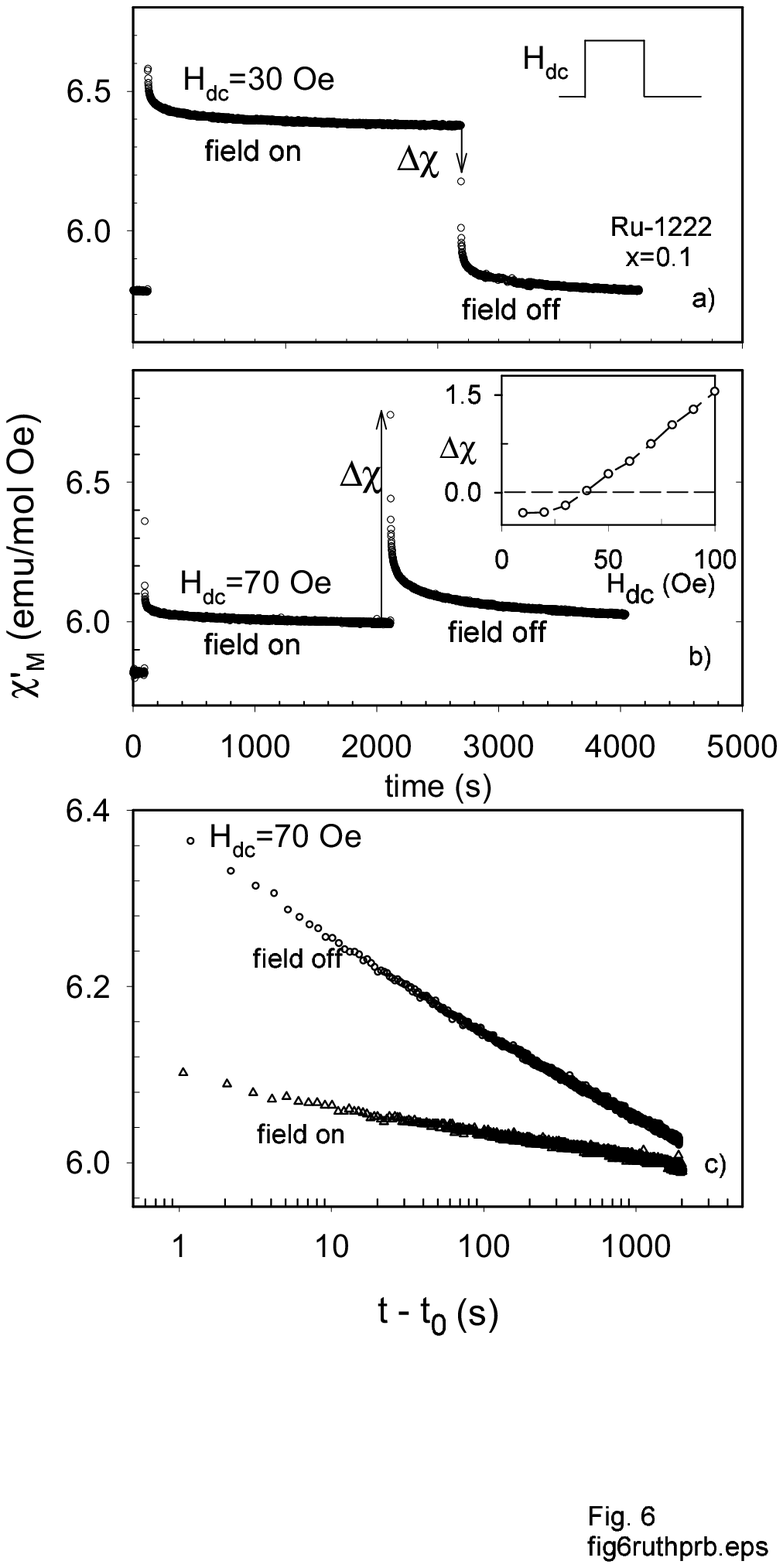,width=7cm,clip=}
\caption{a) Ac susceptibility versus time for Ru-1222, x= 1.0 at 80K,
with $H_{dc}=0Oe$ initially, then $H_{dc}=30Oe$, and finally
$H_{dc}=0Oe$, shown schematically in the Inset. ($\Delta\chi$),
defined as change in susceptibility immediately after H$_{dc}$ is
switched off, is negative.  Note relaxation of susceptibility. b) All
conditions the same as in a) except $H_{dc}=70Oe$. ($\Delta \chi$) is
positive for this value of H$_{dc}$.  Note relaxation of
susceptibility.  Inset:  Overshoot $\Delta \chi$ (in units of
emu/moleOe) at 80K versus H$_{dc}$.  Note change from negative (no
overshoot) to positive (overshoot) at $H_{dc} \approx H_{sf}$. c) Ac
susceptibility versus logarithm of time for Ru-1222, x= 1.0, with
$H_{dc}=70Oe$ and 80K. ($t_{0}$) is time at which H$_{dc}$ is either
switched on or off.  Data for both {\em field on} and {\em field off}
conditions are included.}
\label{f6}
\end{figure}

a) A step-like change of the dc magnetic field causes i) the AC
susceptibility to switch to a new value, and ii) the ZFC equilibrium
state changes to a metastable magnetic state.  The magnitude of the AC
susceptibility change $(\Delta \chi)$ (Fig.\ \ref{f6}a,b) is positive
when H$_{dc}$ is turned on, and can be either positive or negative
when H$_{dc}$ is turned off.  The metastable magnetic state exhibits
logarithmic relaxation, a phenomenon variously ascribed to
disaccommodation~\cite{kro,mur} or magnetic aftereffect~\cite{str}. It
is particularly noteworthy, as Fig.\ \ref{f7} illustrates, that Ru-
1212 samples do not exhibit any indication of AC susceptibility
relaxation;
\begin{figure}[t]
\epsfig{file=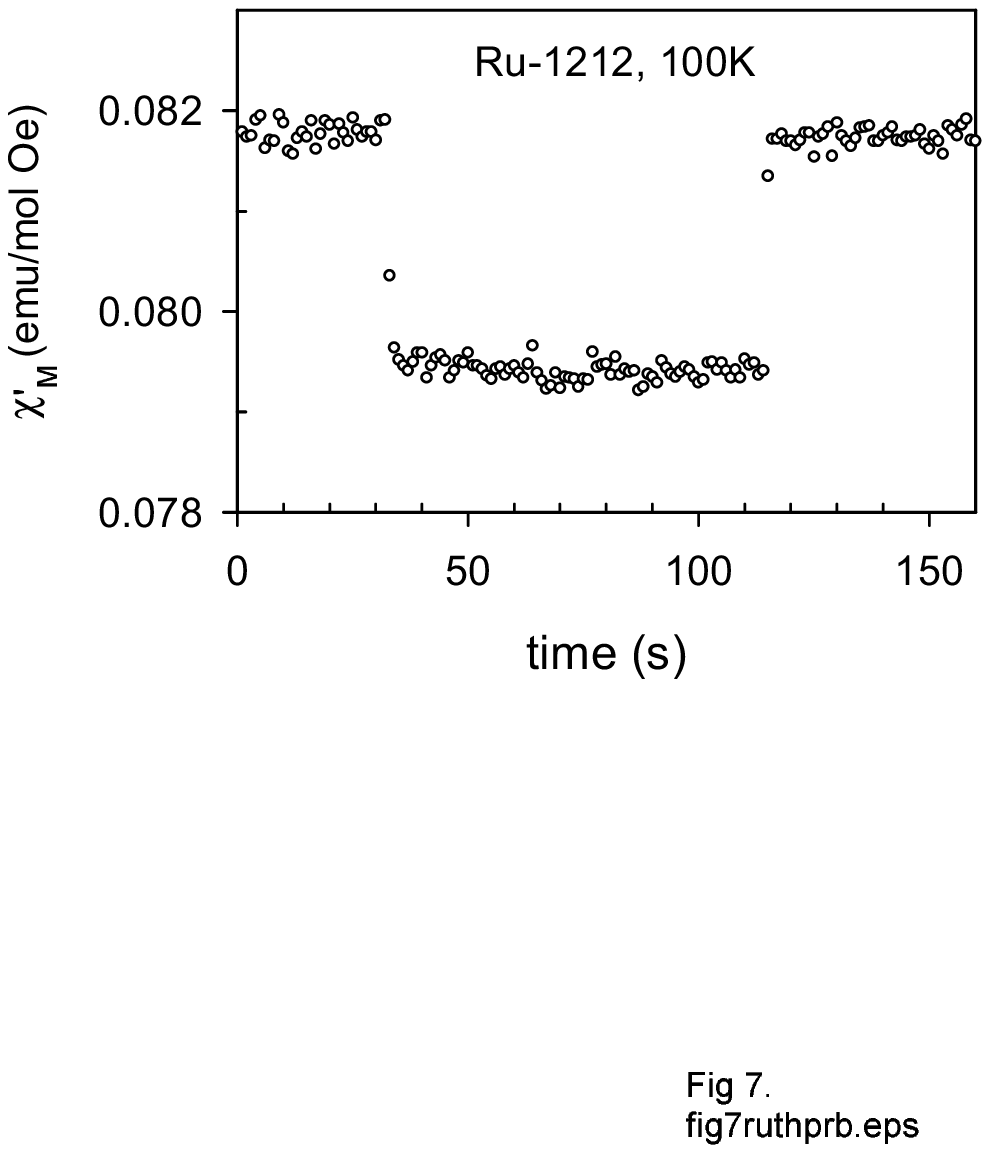,width=7cm,clip=}
\caption{Ac susceptibility versus time for Ru-1212 at 100K in the
applied
rectangular pulse  of magnetic field, as specified in Fig.6a. No time
relaxations can be detected in Ru-1212 sample.}
\label{f7}
\end{figure}

b) The AC susceptibility relaxation is logarithmic in time for both
H$_{dc}$ on and H$_{dc}$ off, and follows the functional form $\chi(t)
=\chi_{0} [1- \alpha ln(t-t_{0})]$. The parameters $\chi_{0}$ and the
relaxation rate $\alpha$ depend on temperature, H$_{dc}$, and the
magnetic history (whether H$_{dc}$ was turned on or off);

c) For $|$H$_{dc}$$|$ above a threshold value of 40 Oe, when
H$_{dc}$ is turned off there is a pronounced `overshoot' phenomenon
with a sizeable positive $(\Delta \chi)$ (see Fig.\ \ref{f6}b);

d) Surprisingly, applying a rectangular field pulse results in a
magnetic state with an {\em increased} AC susceptibility (Fig.\
\ref{f6}). The logarithmic relaxation over several decades of time
incdicates that the excited AC susceptibility persists.

\subsection{AC susceptibility in sweeping magnetic field: Observation of inverted hysteresis}

\begin{figure}[t]
\epsfig{file=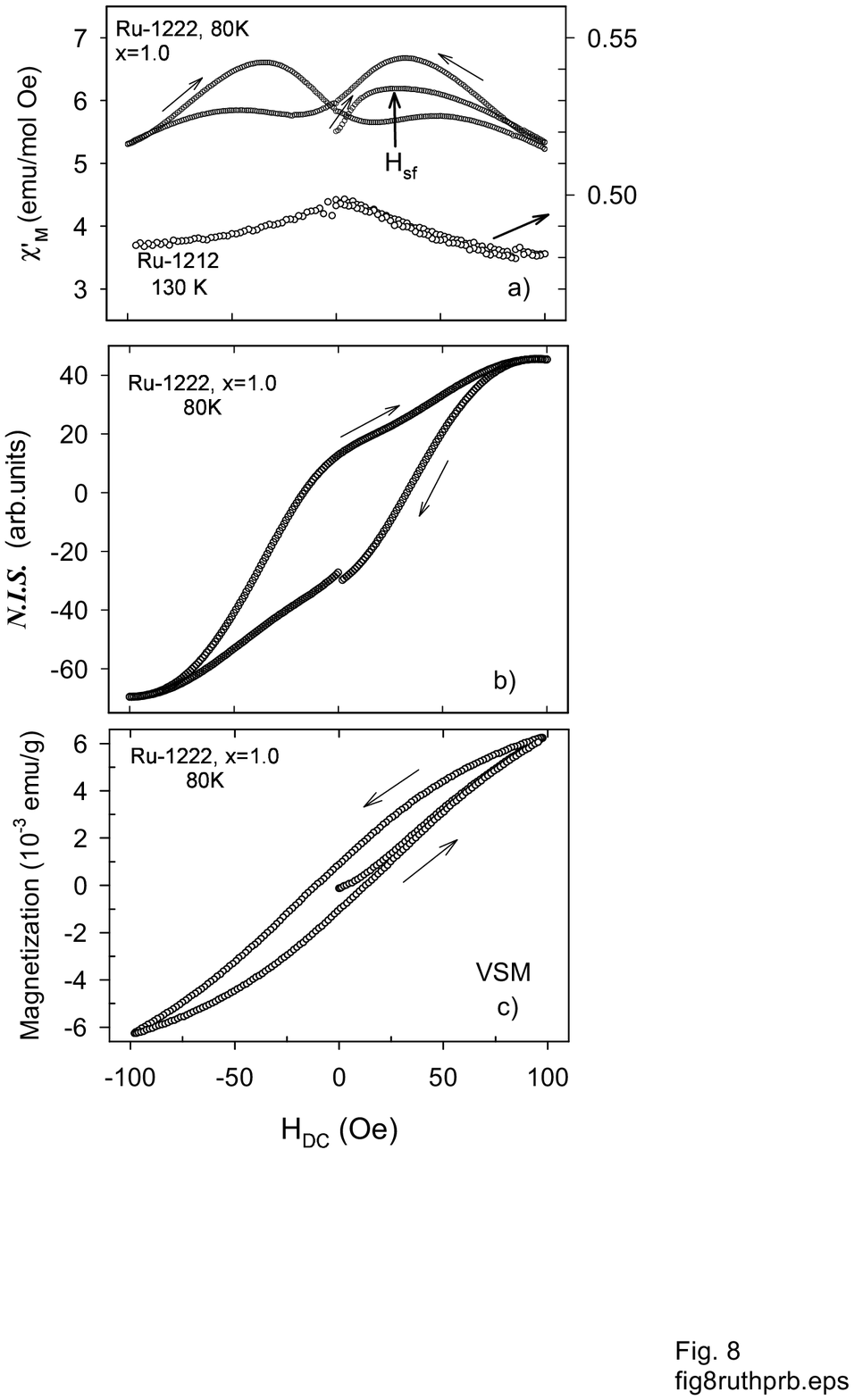,width=7cm,clip=}
\caption[delfsa]{a) Left axis:  `Butterfly' hysteresis for Ru-1222, x=
1.0 at 80K.  Right axis:  Analogous data for Ru-1212, just below
magnetic ordering temperature. Unlike Ru-1222, note for Ru-1212 a
monotonic decrease in ac susceptibility with increasing magnitude of
the dc magnetic field, and no hysteresis. b) Numerical integral $NIS$
($ \equiv \int_{0}^H \chi(h)dh$) of butterfly susceptibility shown in
a) versus H$_{dc}$ for Ru-1222. The units are arbitrary. Note the
inverse hysteresis loop. c) DC (vibrating-sample) magnetization
hysteresis for the same Ru-1222 sample as in a). A normal (counter-
clockwise) circulations is observed.}
\label{f8}
\end{figure}

\begin{figure}
\epsfig{file=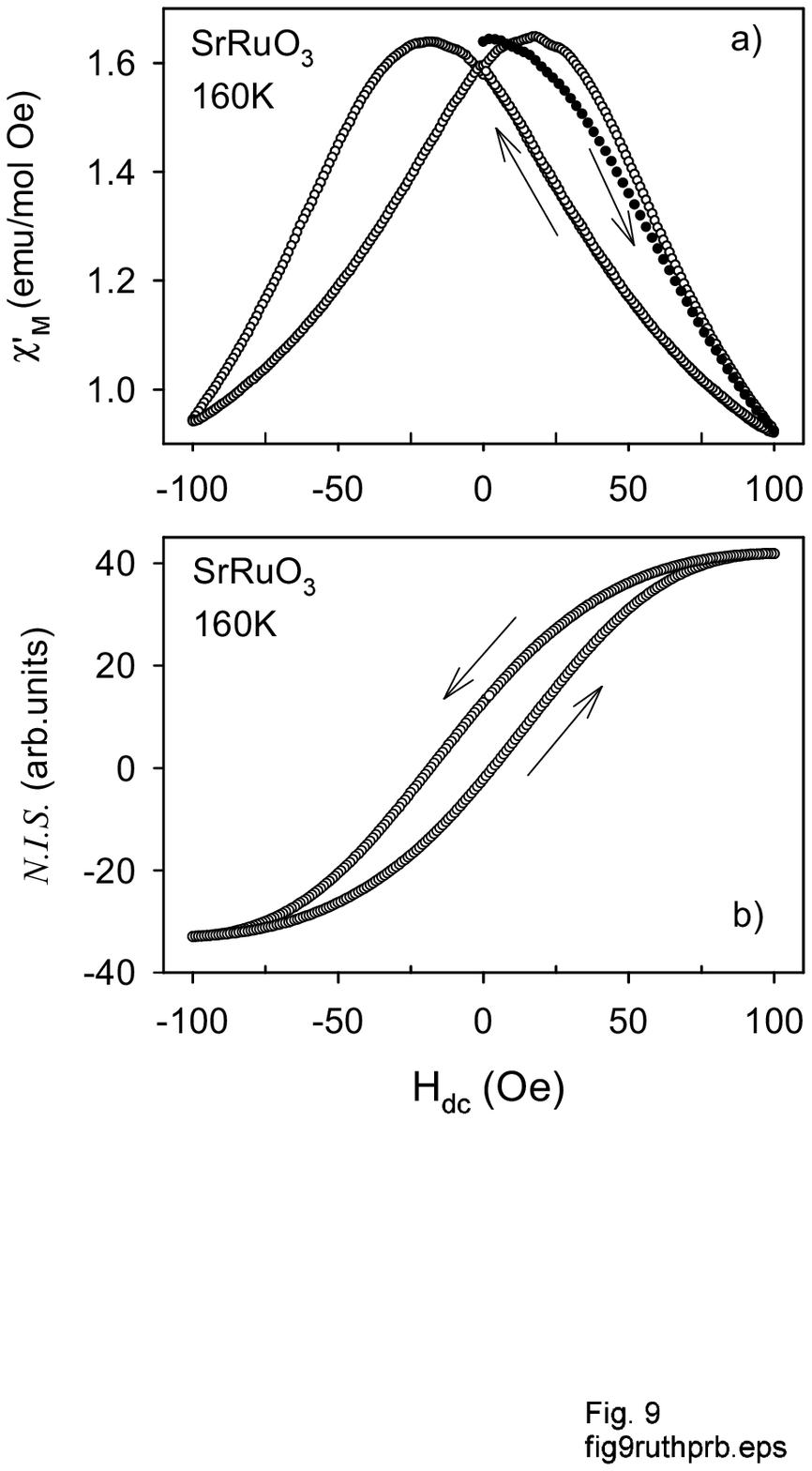,width=7cm,clip=}
\caption{a) Butterfly hysteresis for the reference ferromagnet
SrRuO$_{3}$ at 160K (i.e., just below $T_{c}=165K$, its ordering
temperature). Filled circles designate the virgin hysteresis branch,
characterized by no maximum or other features. Note the response for
increasing and decreasing H$_{dc}$ are opposite to that of Ru-1222. b)
$NIS$ for SrRuO$_{3}$ at 160K.  Note that this hysteresis corresponds,
by all means, to the standard ferromagnetic one.}
\label{f9}
\end{figure}

We also swept the dc magnetic field in an almost continuous fashion,
with increments typically of 1 Oe.  The most striking behavior, as
shown in Fig.\ \ref{f8}, is observing an inverted hysteresis
phenomenon for Ru-1222 that is also entirely absent for Ru-1212 or
SrRuO$_{3}$.
To measure the classic magnetization hysteresis, one ramps the applied
magnetic field (H) from positive to negative and back, and
continuously measures the magnetization M(H).  In a similar,
`butterfly' hysteresis technique~\cite{sal}, the AC susceptibility
$\chi_{ac}(H)$, is measured rather than the magnetization.  Generally,
these two hysteresis loops yield similar information~\cite{fus2}. For
instance, the characteristic maxima in butterfly hysteresis (Figs.\
\ref{f8},\ \ref{f9}) define the coercive field~\cite{sal}. Fig.\
\ref{f8}a shows typical butterfly hysteresis data taken for Ru-1222 and
Ru-1212 samples.  The data establish that the two types of
ruthenocuprates exhibit qualitatively different responses.  There are
also pronounced differences between the Ru-1222 data and the results
for SrRuO$_{3}$ (Fig.\ \ref{f9}a). The most striking difference is the
inverted sense of loop circulation for Ru-1222: the AC susceptibility
signal is consistently larger for the field-decreasing branch compared
to the field-increasing branch.  To our knowledge, this is the first
observation of inverted butterfly loops in a bulk magnetic system. The
numerical integration of the butterfly hysteresis is shown for Ru-1222
(Fig.\ \ref{f8}b) and SrRuO$_{3}$ (Fig.\ \ref{f9}b).
While the results for SrRuO$_{3}$ exhibit the counter-clockwise
pattern of the usual magnetization hysteresis, the integrated
butterfly of the Ru-1222 sample exhibits an inverted (clockwise)
hysteresis loop.  It is noteworthy that the vibrating-sample
magnetometer measurements on the same Ru-1222 (Fig.\ \ref{f8}c) shows
the dc magnetization hysteresis loop with a `normal' (counter-
clockwise) sense of circulation.  Therefore, the inverted hysteresis
phenomenon represents a unique, dynamical feature of the Ru-1222
system, arising from the field-induced and AC magnetic field-assisted
metastable magnetic states observed in Fig.\ \ref{f6}. Other
noteworthy features of the Ru-122 butterfly hysteresis include: i) the
presence of a maximum even in the ZFC (virgin) curve, ii) pronounced
dependence on the observing time used to obtain the data, and iii) the
presence of two- rather than the expected one- maxima per field-
increasing or field-decreasing branch. As Fig.\ \ref{f8}a shows, the
virgin branch exhibits a maximum at a characteristic dc magnetic field
(H$_{sf}$). In simple ferromagnets, the virgin curve typically does
not exhibit a maximum~\cite{fus3} because the remanence, and thus
coercive field~\cite{fus2}, builds up only after the first field
swing, as shown for SrRuO$_{3}$ (Fig.\ \ref{f9}a). The quantitative
size of the butterfly hysteresis loop depends on the observation time,
which is another indication that the metastable magnetic states are
involved.  Qualitatively, though, over the range of sweep times we
studied (one minute to one day), the inverted butterfly loops exhibit
the same features.  Also noteworthy is that the field H$_{sf}$ is
close to the minimum field needed to apply in order to obtain closed
butterfly loops: if the range of sweeping field was narrower than 
($-$H$_{sf}$,$+$H$_{sf}$) no closed loops would be observed whatever.
Instead, the AC susceptibility signal would merely systematically
diminish from cycle to cycle.
\begin{figure}
\epsfig{file=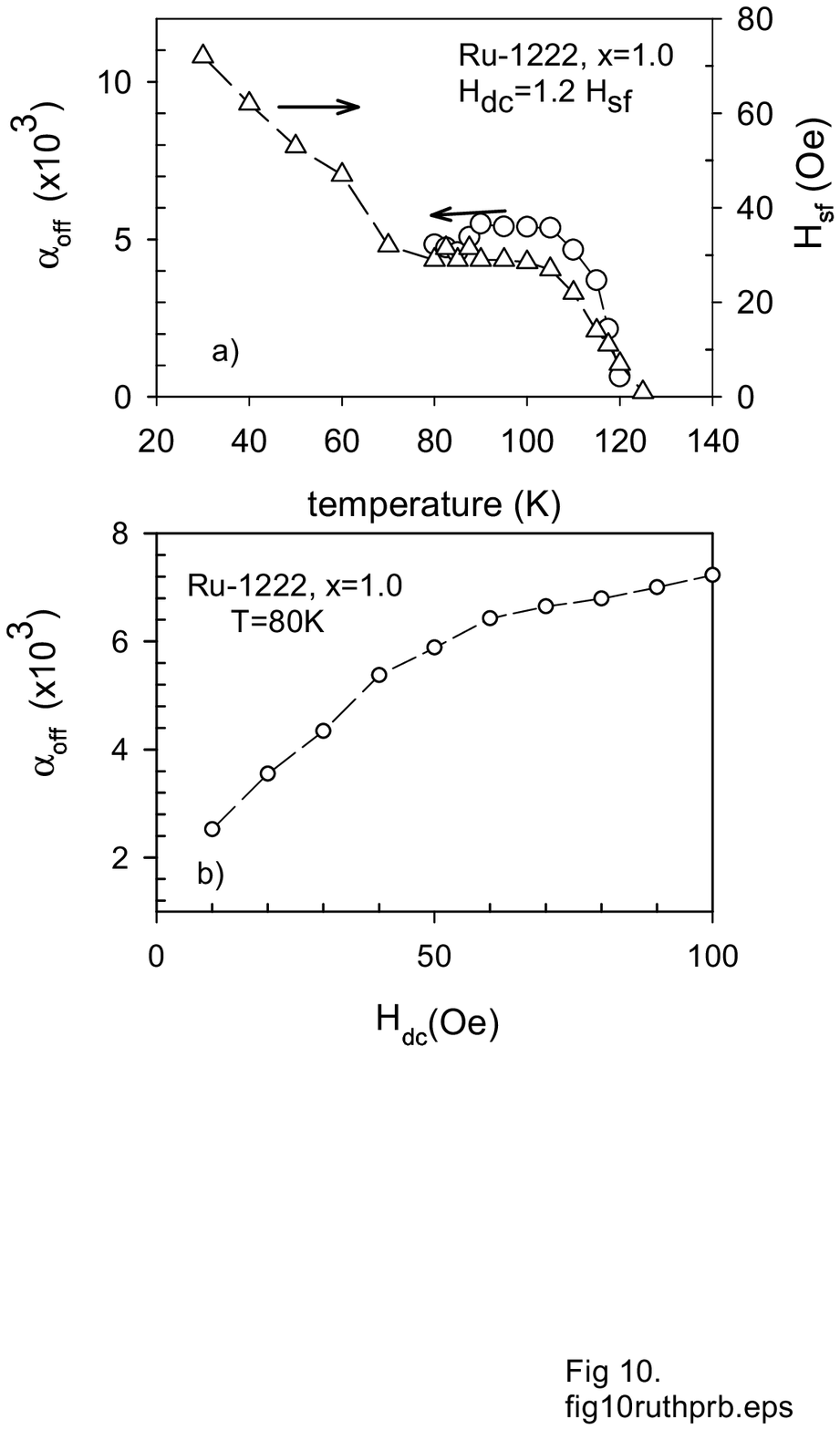,width=7.5cm,clip=}
\caption[delfsa]{a) Left axis: $\alpha_{off}$, the relaxation rate
constant defined in text, when H$_{dc}$ is switched off, versus
temperature.
Right axis: The field attributed to spin flop, H$_{sf}$ (Fig.8a),
versus
temperature.  Note that both quantities are zero at temperatures
above the susceptibility peak. b) $\alpha _{off}$, at 80K, versus
H$_{dc}$.}
\label{f10}
\end{figure}
Fig.\ \ref{f10}a illustrates how H$_{sf}$ and $\alpha_{off}$, the
logarithmic relaxation rate, change with temperature.  The two
parameters-logarithmic relaxation and inverted hysteretic behavior-
exhibit virtually identical temperature dependence, indicating that
the two phenomena have a common origin in Ru-1222.  Another indication
that the two phenomena are interrelated is shown in Fig.\ \ref{f10}b.
$\alpha_{off}$ changes rapidly in small dc magnetic fields, but
saturates at H$_{dc}$ $\geq$ H$_{sf}$. Another commonality is the
change of $\Delta \chi$ with H$_{dc}$ (Fig.\ \ref{f6}b, inset).
$\Delta\chi$ becomes positive above H$_{sf}$ and thereafter
monotonically increases with increasing H$_{dc}$.  Again, these
qualitative changes in relaxation parameters are connected to the
inverse hysteretic behavior. 

\section{discussion} 

The first step to interpreting these results is to determine whether
the results are intrinsic or extrinsic.  The samples are
polycrystalline, so extrinsic sources can include magnetic dynamics of
single domain grains with intergranular magnetic interactions. A
similar question has arisen~\cite{gup,hwa} in studies of
polycrystalline (La,Sr)$_{3}$Mn$_{2}$O$_{7}$.  For our samples, the
microstructure (Fig.\ \ref{f2}) indicates that the phenomena are
intrinsic.  The largest effects were measured on Ru-1222 (x=1.0)
samples having barely detectable grain boundaries with large and
densely packed crystalline grains.  The effects are present in Ru-1222
(x=0.5) but absent in Ru-1222, although these samples have very
similar microstructures with grain size of $1-2 \mu m$ and pronounced
grain boundaries.  We conclude that the phenomena reported in Figs.3-
10 are predominantly intrinsic, due to magnetic interactions within
the unit cell. The non-monotonic ZFC AC susceptibility for different
dc magnetic fields (Fig.\ \ref{f4},\ \ref{f5}) can be naturally
interpreted as indicating the coexistence of antiferromagnetic (AFM)
and ferromagnetic (FM) magnetic ordering in Ru-1222, with the magnetic
order spontaneously occurring below T$_{M}$.  As discussed further
below, we argue that a small dc magnetic field partially cancels the
AFM component, which increases the magnetization of the sample.  This
behavior leads to first an increase in AC susceptibility, with a
decrease as H$_{dc}$ increases further.  A pronounced dependence of AC
susceptibility on the balance between FM and AFM correlations has
recently been reported~\cite{pat} in (La,Sr)$_{3}$Mn$_{2}$O$_{7}$.
Ref.~\onlinecite{pat} reports the onset and growth of AFM
correlations, accompanied by a remarkable drop in the AC
susceptibility, which is consistent with our interpretation.  One
noteworthy difference between this report and Ref.~\onlinecite{pat} is
that in Ru-1222, the AFM contribution is tuned by the dc magnetic
field, while in Ref.~\onlinecite{pat} the AFM correlations are
controlled by varying the stoichiometry.  By contrast to Ru-1222, Ru-
1212 exhibits a monotonic decrease of the AC susceptibility with
increasing dc magnetic field. In Ru-1212, Ref.~\onlinecite{lyn} argues
from neutron scattering data that there is a G-type AFM spontaneous
magnetic order.  We argue that applying a small (0- 100 Oe) dc
magnetic field to Ru-1212 is not large enough to induce any FM order,
while such small fields are sufficient in Ru-1222. The butterfly
hysteresis data provides information about the nature of the AFM
component of magnetic order.  For Ru-1222, the hysteresis loop from AC
susceptibility data is inverted.  A theoretical model for such
inverted hysteresis loops~\cite{aha} indicate that inverted hysteresis
loops can arise in exchange-coupled layered magnetic `sandwiches'
provided that the intralayer coupling is significantly larger than the
interlayer coupling.  The demagnetizing boundary effects, present in
any real finite-size sample, was explicitly taken into account and
shown to be crucial for the model predictions. Ref.~\onlinecite{aha}
also calculated the conditions needed to assure that such inverted
hysteresis loops not to violate the second law of thermodynamics.
Previous experimental reports of inverted hysteresis
loops~\cite{pou,cou} have been limited to magnetic multilayers and
nanoscale magnetic films. Ref.~\onlinecite{pou} argues that in their
samples adjacent layers have magnetic moments with AFM coupling
between adjacent layers.  Ref.~\onlinecite{pou} also demonstrated that
the inverted hysteresis loop behavior disappears in their samples if
the AFM interlayer coupling is changed to a FM interlayer coupling.
The present report, to our knowledge, is the first to show inverted
hysteresis loops for bulk magnetic systems. We argue that the presence
of such inverted hysteresis loops in Ru-1222 arises from RuO$_{2}$
layers with FM magnetic moments, combined with AFM coupling between
the RuO$_{2}$ planes.  A similar conclusion has recently been made for
the layered manganite (La,Sr)$_{3}$Mn$_{2}$O$_{7}$ based on magneto-
optical~\cite{wel} and neutron diffraction~\cite{per} data; in this
compound the MnO$_{2}$ plays the role of the FM layers. 

Observing inverted hysteresis loop behavior provides support for a
magnetic multilayer scenario in Ru-1222.  Arguing for AFM interlayer
coupling, however, requires more specific experimental support. The
most direct support would be evidence of a spin flop transition
through which the net magnetization of adjacent, weakly AFM coupled,
layers increases~\cite{die}.  Unfortunately, there are no single
crystal samples of Ru-1222 available to obtain such direct evidence,
e.g., by neutron diffraction.  We argue, however, that our results
support the presence of a spin flop transition at the characteristic
field H$_{sf}$ in favorably oriented grains of our polycrystalline Ru-
1222 samples.  It is important to note that the butterfly
hysteresis loops we measured would close only for applied fields
larger than H$_{sf}$ , which is consistent with a spin flop transition
and the onset of irreversible behavior.  We attribute the maxima in
the initial AC susceptibility data (Fig.\ \ref{f8}a) to a spin flop
transition.  This assignment is quite similar to a recent report on
AFM ordered chain-ladder compounds~\cite{iso}. In
Ref.~\onlinecite{iso}, the position of the susceptibility peak defines
the spin flop field.  This field is substantially higher in
Ref.~\onlinecite{iso} than in this report due to the different nature
of AFM interactions in the two systems.  However, our experimental
values of a spin flop field below 100 Oe is compatible with the field
values measured in magnetic trilayer and multilayer systems, which are
several order of magnitude smaller than bulk AFM systems~\cite{die}.

\subsection{Model for magnetic coupling in Ru-1222} 

\begin{figure}
\epsfig{file=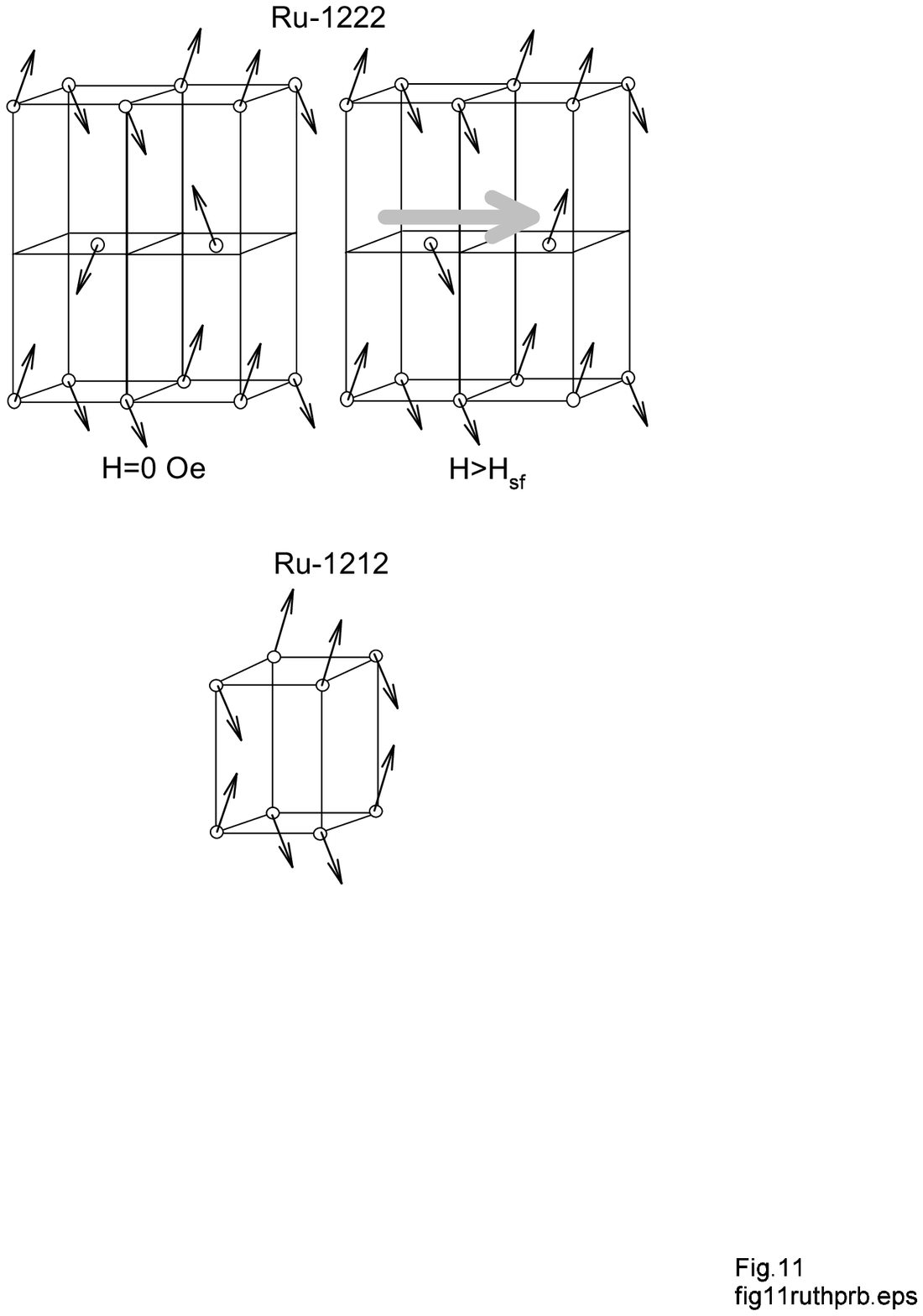,width=8cm,clip=}
\caption{Model for magnetic structure of Ru-1222. The region two unit
cells wide is shown schematically. Circles designate the Ru-ions and
arrows the associated magnetic moments. A widely accepted
model~\cite{jor} for magnetic structure of Ru-1212 is also shown for
comparison. The global order is G-type antiferromagnetic in both
systems. In Ru-1222, small ferromagnetic components-projection of the
moments to the RuO$_{2}$ planes- are antiparallel in H=0. The in-plane
components become mutually parallel - ferromagnetic- by application of
small spin flop field H$_{sf}$. Thick grey arrow designates the field
direction.}
\label{f11}
\end{figure}

While our report does not include any determination of the magnetic
structural order, the results are compatible with a simple model,
shown schematically in Fig.\ \ref{f11}.  We start with the generally
accepted model for the magnetic structure of the more thoroughly
investigated, and simpler, Ru-1212.  It is well
established~\cite{lyn,jor} that the dominant magnetic order is a G-
type antiferromagnetic structure in which the Ru moments are aligned
antiparallel in all crystallographic directions.  The details of the
magnetic order, and the stability of a particular ground
state~\cite{nak}  can be interpreted only by explicitly including
rotations and tilting of the RuO$_{6}$ octahedra and considering the
orientation of the magnetic moments. A weak ferromagnetism originates
from canting of the Ru moments~\cite{fel1}. The canting arises from
the Dzyaloshinsky-Moriya~\cite{mor} antisymmetric superexchange
interaction which, by symmetry, follows from the fact that the
RuO$_{6}$ octahedra tilt away from the crystallographic c direction;
there is still a controversy as to whether the tilting around the axis
perpendicular to the c-axis is actually observed~\cite{wil1,jor,mcl}.
In Ru-1212 samples containing magnetic (Gd) ions, the dipolar field of
the in-plane ferromagnetic components induces an additional
ferromagnetic component~\cite{wil,jor}.  Very recently, results of a
structural investigation of the Ru-1222 compound (x=1.0) indicates
that there are no important differences in rotation or tilting angles
between the Ru-1212 and Ru-1222 (x=1.0) compounds~\cite{wil1}, in
spite of Ru-1212 exhibiting dominant antiferromagnetic and Ru-1222
dominant ferromagnetic spontaneous magnetic order. We suggest that the
magnetic ordering in both compounds is a variation of the G-type
antiferromagnetism, while variation in dc magnetic properties,
particularly in low fields, arises from small differences in the Ru-Ru
interaction within their respective unit cells.  The unit cells are
different: in Ru-1222 there is a structural phase shift of half of the
RuO$_{2}$ planes that leads to an approximate doubling of the unit
cell.  The phase shift arises from the presence of the fluorite-
structure block Eu$_{2-x}$Ce$_{x}$O$_{2}$ replacing the rare earth ion
in Ru-1212.  Thus, in Ru-1222, nearest neighbor (Ru) ions are not
vertically aligned, while they are in Ru-1212.  This difference in
structure naturally leads, in Ru-1222, to having the relative
alignment of the in-plane components in adjacent RuO$_{2}$ planes
antiparallel, which is energetically favored by a bare dipole-dipole
interaction.  Fig.\ \ref{f11} shows the magnetization tilting scheme
we propose for Ru-1222.  Our experimental results fit quite naturally
with such a picture: since the dipole interaction is very weak- the
energy needed to reverse the in-plane Ru moment component is small- a
small applied magnetic field can easily transform the spontaneous AF
order, via a spin flop mechanism, into a ferromagnetic orientation.
This is exactly what our measurements indicate.  Our model is, apart
from the unusual dipole-dipole interaction, the same as the models
used to explain weakly AF coupled magnetic multilayers~\cite{die,pri}.
We note that our model does not take into account the role of the
(Ce), which is expected to have a non-zero magnetic moment.  Our model
is, however, of use in understanding the qualitative features of the
Ru-1222 experimental data.

The most unusual feature of our model is the pronounced role of the
dipole-dipole interaction; various magnetic exchange interactions
(e.g., superexchange, double exchange) are more commonly employed to
explain magnetic coupling.  We argue that a dipole-dipole interaction
makes sense because the nearest RuO$_{2}$ layers are far apart, with
insulating and non-magnetic layers in between.  Experimentally, the
logarithmic time dependence of the magnetic relaxation (Fig.\
\ref{f6}c) argues for the long-range dipole-dipole interaction; it is
well known~\cite{lot} that such a long-range interaction can account
for a logarithmic relaxation behavior without further assumptions.  We
also considered the possibility that the logarithmic relaxation
behavior might be due to domain-wall stabilization (disaccommodation)
involving a broad range of activation energies, as has been recently
applied to data in a perovskite manganite~\cite{mur1,mur}.  However,
one of us (IF) and colleagues have performed temperature-dependent x-
ray diffraction studies of Ru-1222.  The measurements indicate no
structural change with temperature- such changes are necessary for
disaccommodation~\cite{fel3}. Thus both the weak AF coupling and the
logarithmic susceptibility relaxation support our model.

\section{conclusion}

In summary, we have presented data indicating that Ru-1222 exhibits
qualitatively new magnetic behavior, including magnetic logarithmic
relaxation, inverted hysteresis loops, and metastable magnetic
states.  None of these behaviors are observed in Ru-1212.  Our
results are interpreted within a model for the magnetic structure for
Ru-1222 that, assuming a G-type AF global magnetic order known to
describe Ru-1212, attributes the interlayer magnetic coupling to a
dipole-dipole interaction.  We interpret the hysteretic behavior as
similar to that reported for magnetic multilayers, trilayers, and
some manganites, due to spin flop transitions converting the
spontaneous (H = 0) AF order between components into a ferromagnetic
order.

\acknowledgments  

We benefited from conversations with Robert Joynt.  We also thank K.
Zadro for the VSM measurements.  Financial support was partially
provided by the U.S.- Israel Binational Science Foundation, the U.S.
D.O.E.,  the SCOPES program of the Swiss National Science Foundation, 
Fonds National Suisse de la Recherche Scientifique and EPFL.

{\em Note added}: After completion of this work and after the original
version of the manuscript was submitted in short form~\cite{ziv}, we
became aware of two reports that are related to our report, including
the work by Ohkoshi et.al.~\cite{ohk} on spin flip transitions in bulk
materials, and by Y.Y. Xue et.al.~\cite{xue} on
RuSr$_{2}$(Gd,Ce)$_{2}$Cu$_{2}$O$_{10-y}$ using magnetization.



\begin{thebibliography}{99}
\bibitem{fus2}The equivalency of butterfly and magnetization
hysteresis is simple consequence of the relationship:  
$M(H)=\int_{0}^H \chi(h)dh$ valid - up
to an integration constant- in simple isotropic ferromagnets. 
However, in complex magnetic systems replacing $\chi$ (a tensor) 
by a scalar is not approved and the relationship does not hold.
\bibitem{fus3}In principle, the observed maximum could be ascribed 
to a cross over between reversible and irreversible
dynamics of ferromagnetic domains. However, the idea of
reversibility, at least in its elementary form, is incompatible with 
logarithmic relaxations characterizing the whole 0-100 Oe range.  
Therefore, it seems that the latter interpretation cannot be applied 
in its original transparent form.
\end{thebibliography}
\end{document}